\documentclass[conference,a4paper]{APSIPA2026}

\usepackage{amsmath}
\usepackage{graphicx}
\usepackage{multirow}
\usepackage{threeparttable}
\usepackage[backend=biber,style=ieee,]{biblatex}
\usepackage{graphicx}
\usepackage{booktabs}
\usepackage{multirow}

\usepackage{xcolor}
\usepackage{tikz}
\usepackage{pifont}

\newcommand{\cmark}{\ding{51}}
\newcommand{\xmark}{\ding{55}}

\newcommand{\greencheck}{{\cmark}}
\newcommand{\redcross}{{\xmark}}

\usepackage{geometry}
\usepackage{fancyhdr}

\fancypagestyle{firststyle}{
  \fancyhf{}
  \fancyhead[C]{2026 Asia Pacific Signal and Information Processing Association Annual Summit and Conference (APSIPA ASC)}
}

\begin{document}

\title{AudioTurbo: Fast Text-to-Audio Generation with Rectified Diffusion}

% \author{
% \authorblockN{
% Author 1\authorrefmark{1} and
% Author 2\authorrefmark{2}
% }

% \authorblockA{
% \authorrefmark{1}
% Vietnam National University, Hanoi, Vietnam \\
% E-mail: xx@vnu.edu.vn  Tel/Fax: +84-XXXXXXXX}

% \authorblockA{
% \authorrefmark{2}
% Hanoi University of Science and Technology, Vietnam \\
% E-mail: yy@hust.edu.vn  Tel/Fax: +84-XXXXXXXX}
% }

\author{
\authorblockN{
Junqi Zhao\authorrefmark{1},
Jinzheng Zhao\authorrefmark{1},
Haohe Liu\authorrefmark{1},
Yun Chen\authorrefmark{2},
Lu Han\authorrefmark{3},
Haoyang Zhang\authorrefmark{4},
\\Xubo Liu\authorrefmark{1},
Mark Plumbley\authorrefmark{2}, and
Wenwu Wang\authorrefmark{1}
}

\authorblockA{
\authorrefmark{1}
Centre for Vision, Speech and Signal Processing (CVSSP), University of Surrey, UK\\
E-mail: junqi.zhao@surrey.ac.uk, W.Wang@surrey.ac.uk
}

\authorblockA{
\authorrefmark{2}
King's College London, UK
}

\authorblockA{
\authorrefmark{3}
Institute of Acoustics, Chinese Academy of Sciences, China
}

\authorblockA{
\authorrefmark{4}
Peking University, China
}
}

\maketitle
\thispagestyle{firststyle}
\pagestyle{empty}

\begin{abstract}
  % Diffusion models have significantly improved the quality and diversity of audio generation but are hindered by slow inference speed. Rectified flow enhances inference speed by learning straight-line ordinary differential equation (ODE) paths. However, this approach requires training a flow-matching model from scratch and tends to perform suboptimally, or even poorly, at low step counts. To address the limitations of rectified flow while leveraging the advantages of advanced pre-trained diffusion models, this study integrates pre-trained models with the rectified diffusion method to improve the efficiency of text-to-audio (TTA) generation. Specifically, we propose AudioTurbo, which learns first-order ODE paths from deterministic noise-sample pairs generated by a pre-trained TTA model. Experiments show that our approach outperforms a flow-matching-based acceleration model in both generation quality and inference speed.
  Diffusion models have significantly improved the quality and diversity of text-to-audio (TTA) generation but are hindered by slow inference speed. Recent flow-based approaches have been explored for accelerating TTA generation. However, these approaches typically rely on dedicated flow-based generative frameworks and do not directly exploit existing pre-trained diffusion-based TTA models. To address this limitation while leveraging the strong generative capabilities of pre-trained diffusion models, we integrate rectified diffusion into these models to improve TTA generation efficiency. Specifically, we propose AudioTurbo, which learns first-order approximate ordinary differential equation (ODE) paths from deterministic noise-sample pairs generated by a pre-trained TTA model. Experiments show that our approach outperforms a flow-matching-based acceleration model in both generation quality and inference speed.
\end{abstract}

\begin{IEEEkeywords}
  Audio generation, diffusion model, flow matching, rectified diffusion, text-to-audio. 
\end{IEEEkeywords}

\section{Introduction}

\begin{figure*}[tbp]
  \centering
  \includegraphics[width=0.8\linewidth]{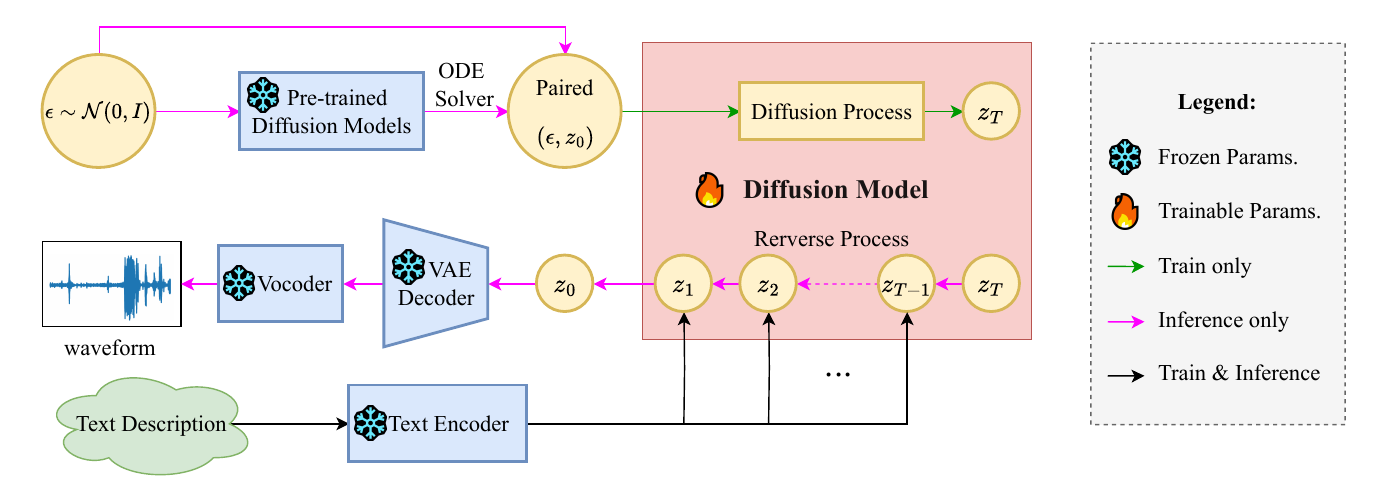}
  \caption{An overview of AudioTurbo architecture. Note that the trainable parameters are initialized using the pretrained TTA model, Auffusion.}
  \label{fig:speech_production}
\end{figure*}

Text-to-audio (TTA) generation is a task that aims to generate audio samples based on natural language descriptions \cite{kreukaudiogen, liu2023audioldm, ghosal2023text}. In previous research, TTA methods are typically classified into two main categories: language model-based methods and diffusion model-based methods. The first category quantizes audio into discrete tokens and uses either autoregressive (AR) or non-autoregressive (NAR) transformer architectures for modeling. AudioGen \cite{kreukaudiogen} is a notable AR generative model that leverages learned discrete audio tokens and a transformer decoder to synthesize audio based on text descriptions. UniAudio \cite{yang2023uniaudio} leverages large language models (LLMs) to generate  a variety of audio types conditioned on diverse input modalities. The second category applies diffusion models to continuous or discrete representations provided by autoencoders to synthesize audio. Diffsound \cite{yang2023diffsound} generates discrete codes by quantizing audio mel-spectrograms using a vector quantized variational autoencoder (VQ-VAE) \cite{van2017neural}. Inspired by stable diffusion \cite{rombach2022high}, AudioLDM \cite{liu2023audioldm} is the first to employ a continuous latent diffusion model (LDM) to enhance the computational efficiency of diffusion models while maintaining their quality and versatility. To further enhance text understanding beyond AudioLDM and facilitate the learning of complex concepts in textual descriptions, Tango \cite{ghosal2023text} proposes using an instruction-tuned LLM (FLAN-T5) as a text encoder instead of the contrastive language-audio pretraining (CLAP) model used by AudioLDM. Recently, Auffusion \cite{xue2024auffusion} tailors text-to-image LDM frameworks for the TTA task, effectively harnessing their inherent generative capabilities and accurate cross-modal alignment. Like Auffusion, other TTA models~\cite{liu2024audioldm,huang2023make,huang2023make2} also utilize latent variable generation in conjunction with a pre-trained VAE for audio synthesis, delivering outstanding generation quality. These TTA models trained at scale on diverse text-audio corpora can serve as audio foundation models (AFMs) due to their ability to learn transferable representations that support a wide range of downstream tasks, including controllable audio generation \cite{xie2024picoaudio, zhu2026audio}, personalization \cite{yuan2025dreamaudio}, and audio editing \cite{xu2024prompt, hou2025editing}.

Although diffusion-based TTA models achieve high generation quality and diversity through large-scale pre-training, the iterative sampling process of pre-trained latent diffusion models (LDMs) imposes substantial computational costs, resulting in slow inference speeds and limited real-time applicability. To date, several methods \cite{mehta2024matcha, ye2024flashspeech} that achieve fast sampling while maintaining satisfactory generation quality have been proposed for text-to-speech (TTS). However, these methods require training a new model from scratch, which demands large-scale training data and significant computational overhead, and cannot inherit the general knowledge encoded in pre-trained AFMs. 

In the broader domain of TTA generation, several acceleration paradigms have recently been explored. Flow-matching-based approaches, such as LAFMA \cite{guan2024lafma}, apply flow matching in the audio latent space, enabling high-quality generation with substantially fewer inference steps. Consistency-based approaches have also been investigated. ConsistencyTTA \cite{bai2024consistencytta} introduces a classifier-free guidance (CFG)-aware latent consistency model that incorporates CFG into consistency training, requiring only a single network query per generation. SoundCTM \cite{saito2024soundctm} adapts consistency trajectory modeling to sound generation, supporting one-step generation and further quality refinement through deterministic multi-step sampling. AudioLCM \cite{liu2024audiolcm} proposes guided latent consistency distillation with a multi-step ODE solver, achieving high-quality audio generation with only a few inference steps. Another line of research explores rectified flow for accelerating TTA generation. FlashAudio \cite{liu2025flashaudio} introduces a rectified-flow-based framework with specialized designs for timestep sampling, noise--data assignment, and CFG optimization, targeting high-quality one-step generation.

% In the broader domain of general sound generation with TTA, such acceleration methods are still relatively limited. Recently, Guan et al. introduced LAFMA \cite{guan2024lafma}, a model that incorporates flow matching \cite{lipmanflow} into the audio latent space to enable text-guided audio generation. LAFMA is capable of producing high-quality audio samples by utilizing a numerical ordinary differential equation (ODE) solver, effectively reducing the inference steps to around ten. A drawback of LAFMA is its reliance on flow matching, which prevents it from fully utilizing well-performing pre-trained diffusion-based TTA models and potentially achieving superior performance in fewer than $10$ steps.

% a flow-matching-based diffusion process
% However, in the broader domain of general sound generation (TTA), such acceleration methods are still relatively limited.
% During the post-training phase, improving sampling efficiency through approaches such as distillation has become a central focus in developing diffusion-based audio generation models.

% % To directly exploit the strong capabilities of AFMs to accelerate sampling in the post-training stage, we propose AudioTurbo, which integrates rectified diffusion into pre-trained TTA models to enable efficient text-guided audio generation. 

% To directly exploit the strong capabilities of AFMs and accelerate sampling on top of pre-trained models, we propose AudioTurbo, a post-training distillation framework that integrates rectified diffusion into these models to enable efficient text-guided audio generation. 
In this work, we explore a different acceleration paradigm that directly leverages existing pre-trained diffusion-based TTA models while retaining their learned generative capabilities. To this end, we propose AudioTurbo, a post-training distillation framework that integrates rectified diffusion into pre-trained TTA models to enable efficient text-guided audio generation. Rectified diffusion \cite{wang2024rectified} extends the design space of rectified flow to general diffusion models. It is a progressive retraining approach that transforms the random coupling of noise and real data used in diffusion training into a deterministic coupling of pre-sampled noise and generated data. Consequently, the model's predictions maintain consistency throughout the ODE trajectory and continue to follow the same trajectory after each inference step. To our knowledge, this is the first work to investigate rectified diffusion as a post-training framework for accelerating pre-trained diffusion-based TTA models. We conduct experiments using Auffusion as the primary pre-trained TTA model and further evaluate the generalizability of our method using AudioLDM. We also investigate the application of classifier-free guidance to AudioTurbo. Our experiments demonstrate that AudioTurbo maintains competitive generation performance while substantially reducing the number of sampling steps. Demos are available at: \url{https://audioturbo.github.io/demos}.

% To our knowledge, this is the first work to introduce rectified diffusion in the field of audio generation for acceleration. We conduct experiments with Auffusion, a pre-trained diffusion model for TTA, and further investigate the application of classifier-free guidance for AudioTurbo, achieving enhanced audio generation performance. Our experiments demonstrate that AudioTurbo achieves superior performance with significantly fewer sampling steps. Demos are available at: \url{https://audioturbo.github.io/demos}.

%not only utilizes SOTA pre-trained diffusion-based models but also

% To the best of our knowledge, we are the first to introduce rectified diffusion in the field of sound. In addition, we conduct experiments with Auffusion, a state-of-the-art pre-trained diffusion model for TTA, and further investigate the application of classifier-free guidance \cite{ho2022classifier} to direct AudioTurbo, achieving enhanced audio generation performance. Our experiments demonstrate that AudioTurbo can not only utilize state-of-the-art (SOTA) pre-trained diffusion-based models without requiring training from scratch but also achieve superior performance in just a few steps. Specifically, AudioTurbo generates higher-quality audio in only 5 steps while exhibiting better text-audio alignment capabilities compared to the advanced acceleration model LAFMA \cite{guan2024lafma}.

\section{Rectified Diffusion}

% \subsection{Layout}

Diffusion models operate by gradually adding noise to transform a complex data distribution into a known prior distribution and then learning to reverse the process by progressively denoising the prior to reconstruct the original data distribution. The forward diffusion process can be modeled by a continuous-time stochastic differential equation (SDE) \cite{song2020score}:
\begin{align}
  d\mathbf{x}_{t} = f(\mathbf{x}_{t}, t) \, dt + g(t) \, d\mathbf{w}.
  \label{equation:sde}
\end{align}
Here, \( t \) is a continuous time variable with \( t \in [0, T] \). The functions \( f(\cdot) \) and \( g(\cdot) \) denote the drift coefficient and diffusion coefficient, respectively, while \( \mathbf{w} \) represents Brownian motion.

The reverse process has an equivalent deterministic process whose trajectories share the same marginal probability densities as those of the corresponding SDE. This deterministic process is governed by an ODE, referred to as the probability flow ODE \cite{song2020score}:
\begin{align}
  d\mathbf{x}_{t} = \left[ f(\mathbf{x}_{t}, t) - \frac{1}{2} g(t)^2 \nabla_{\mathbf{x}_{t}} \log p_t(\mathbf{x}_{t}) \right] dt,
  \label{equation:ode}
\end{align}
where \( \nabla_{\mathbf{x}_{t}} \log p_t(\mathbf{x}_{t}) \) is the score function. Numerical simulation of ODEs is generally simpler and faster compared to SDEs, making ODEs the preferred choice in most cases.

By using a noise prediction model \( \boldsymbol{\epsilon_\theta}(\mathbf{x}_{t}, t) \) to approximate the score function, Song et al. \cite{song2020score} defined the following diffusion ODE:
\begin{align}
  d\mathbf{x}_{t} = \left[ f(\mathbf{x}_{t}, t) + \frac{g(t)^2}{2\sigma_t} \boldsymbol{\epsilon_\theta}(\mathbf{x}_{t}, t) \right]dt,
  \label{equation:diffusion_ode}
\end{align}
where \( \quad \mathbf{x}_{T} \sim \mathcal{N}(\mathbf{0}, \sigma^2 \mathbf{I}) \). Moreover, Lu et al. \cite{lu2022dpm} provided an exact solution for the diffusion ODE:
\begin{align}
  \mathbf{x}_{t} = \frac{\alpha_t}{\alpha_s} \mathbf{x}_{s} - \alpha_t \int_{\lambda_s}^{\lambda_t} e^{-\lambda} \boldsymbol{\epsilon_\theta}({\mathbf{x}_{t_\lambda(\lambda)}}, t_\lambda(\lambda)) \, d\lambda.
  \label{equation:ode_solution}
\end{align}
The function \( \lambda_t  = \log\left(\frac{\alpha_t}{\sigma_t}\right) \), in which \( \lambda_t = \lambda(t) \), is strictly decreasing with respect to \( t \). Therefore, it has an inverse function \( t_\lambda \). Starting from \( \mathbf{x}_{s} \) at time \( s \), Equation (\ref{equation:ode_solution}) shows that solving for the value at time \( t \) is equivalent to approximating the exponentially weighted integral of \( \boldsymbol{\epsilon_\theta} \) over the range from \( \lambda_s \) to \( \lambda_t \).

% The solution to the diffusion ODE is equivalent to the following equation:
For a first-order approximate ODE path, Equation (\ref{equation:ode_solution}) can be equivalently expressed as: 
% The first-order ODE means that Equation (\ref{equation:ode_solution}) is equivalent to the following equation:
%
\begin{align}
\mathbf{x}_{t}=\frac{\alpha_{t}}{\alpha_{s}} \mathbf{x}_{s}-\alpha_{t} \boldsymbol{\epsilon_{\theta}}\left(\mathbf{x}_{s}, s\right)\left(\frac{\sigma_{s}}{\alpha_{s}}-\frac{\sigma_{t}}{\alpha_{t}}\right)
  \label{equation:ode_solution_equiv}
\end{align}
if and only if \( \boldsymbol{\epsilon_\theta}(\mathbf{x}_t, t) \) remains constant along the same ODE trajectory \cite{wang2024rectified}. By substituting \( s = 0 \), \( \alpha_0 = 1 \), \( \sigma_0 = 0 \), and \( \boldsymbol{\epsilon_\theta}(\mathbf{x}_s, s) = \boldsymbol{\epsilon} \) into Equation (\ref{equation:ode_solution_equiv}), we obtain:
\begin{align}
  \mathbf{x}_{t} = \alpha_t \mathbf{x}_{0} + \sigma_t \boldsymbol{\epsilon}.
  \label{equation:diffusion_form}
\end{align}
This follows the same general form as forward diffusion process, but here, since the epsilon prediction is constant, the noisy data becomes a weighted interpolation of noise and samples that constitute a deterministic pair. This differs from standard diffusion model training, where noise and samples are paired randomly. If perfect coupling of deterministic noise-sample pairs is achieved during training, the model can make consistent predictions along a single trajectory. Theoretically, this enables the same generation results to be obtained with any number of inference steps.

\section{Proposed Method}

We propose AudioTurbo, a generative model based on rectified diffusion for efficient TTA generation. AudioTurbo consists of four key components: 1) a text encoder for generating text embeddings; 2) an LDM based on rectified diffusion for predicting audio latent representations; 3) a VAE decoder that reconstructs the mel-spectrograms; and 4) a vocoder for generating audio samples. The model architecture and the mechanism of rectified diffusion are shown in Figure~\ref{fig:speech_production}.

% 3) a VAE with an encoder that compresses mel-spectrograms into a latent space and a decoder that reconstructs them;

\subsection{Text Encoder}\label{sec:figures}

Unlike AudioLDM \cite{liu2023audioldm}, which utilizes a CLAP \cite{wu2023large} encoder, or Tango \cite{ghosal2023text}, which employs a FLAN-T5 model to encode text descriptions into text embeddings, our experiments use a Contrastive Language-Image Pretraining (CLIP) \cite{radford2021learning} encoder, inspired by the experimental results from Auffusion \cite{xue2024auffusion}, where CLIP achieved a higher Inception Score than CLAP or FLAN-T5 in TTA.

% Motivated by the experimental results from Auffusion \cite{xue2024auffusion}, where CLIP demonstrated a higher Inception Score than CLAP or FLAN-T5 in TTA, which means using CLIP as the text encoder can  achieve better sampling quality and diversity.
% original table
\begin{table*}[ht!]
\centering
% \normalsize
\small
\caption{The comparison between our AudioTurbo and baseline TTA models on the AudioCaps dataset.}
\resizebox{16cm}{!}
% \resizebox{\textwidth}{!}
{
\begin{tabular}{ccccc|cccc|cc}
\toprule
\multirow{2}{*}{\textbf{Model}} &
\multirow{2}{*}{\textbf{Params}} &
\multirow{2}{*}{\textbf{Steps}} &
\multirow{2}{*}{\textbf{RTF}} & \multirow{2}{*}{\textbf{Text}} & 
% \multirow{2}{*}{\textbf{Audio}} & 
% \multirow{2}{*}{\textbf{Params}} &
\multicolumn{4}{c|}{\textbf{Objective Metrics}} & \multicolumn{2}{c}{\textbf{Subjective Metrics}} \\ %\cline{5-9}
& & & & & FD~$\downarrow$ & KL~$\downarrow$ & IS~$\uparrow$ & CLAP (\%)~$\uparrow$ & OVL~$\uparrow$ & REL~$\uparrow$ \\
\midrule
Ground Truth & $-$ & $-$ & $-$ & $-$ & $-$ & $-$ & $-$ & $-$ & $91.30$ & $92.03$ \\
\midrule

Tango & $866$M & $200$ & $1.789$ & 
\greencheck  & $21.97$   & $1.52$ & $7.37$ & $23.6$ & $80.87$ & ${82.30}$ \\
AudioLDM2 & $1.1$B & $200$ & 2.070 & 
\redcross & $36.95$   & $1.82$ & $6.87$ & $22.3$ & $79.63$ & $75.97$ \\
% \midrule
LAFMA & $272$M & $200$ & $1.009$ & 
\greencheck & $29.59$   & $1.56$ & $7.29$ & $22.9$ & $78.00$ & $78.73$ \\
\midrule

% \midrule
% AudioTurbo & $1.1$B & $3$ & AC captions & 
% \greencheck & $29.96$   & $1.52$ & $6.70$ & $25.4$ & $-$ & $-$ \\
AudioTurbo & $1.1$B & $5$ & $0.093$ & 
\greencheck & $22.18$   & $1.30$ & $8.88$ & $29.2$ & $-$ & $-$ \\
AudioTurbo & $1.1$B & $10$ & $0.112$ & 
\greencheck & $\mathbf{20.65}$   & $\mathbf{1.29}$ & $\mathbf{9.40}$ & $\mathbf{29.8}$ & $\mathbf{82.25}$ & $\mathbf{85.58}$ \\
\bottomrule
\end{tabular}
}
\label{tab:AudioCapsResults}
\end{table*}

\subsection{Rectified Diffusion-based LDM}\label{sec:LDM}

The goal of the LDM is to reconstruct the audio representation \( z_0 \) in the latent space, conditioned on the text embedding \( \tau \) provided by a pre-trained text encoder. This is accomplished by reversing a forward diffusion process that progressively transforms the clean data distribution into a standard normal distribution, following a predefined noise schedule,
\begin{align}
    q\left(z_{t} \mid z_{0}\right)
    =\mathcal{N}\left(z_t;~{\alpha}_{t} z_{0},{\sigma}_{t}^{2} I\right), 
    \label{equation:forward_process}
\end{align}
this forward process demonstrates that any latent variable \( z_t \) can be directly sampled from \( z_0 \), and it represents a weighted interpolation of clean data and Gaussian noise. The generative (reverse) diffusion process can be conditioned on text embeddings by employing a deep neural network (DNN), \( \epsilon_\theta(z_t, t, \tau) \), to predict noise, thereby reconstructing \( z_0 \). The loss function for the optimization process is given as follows:
\begin{align}
    L(\theta) = || {\epsilon} - \epsilon_\theta(z_t, t, \tau) ||^2,
    \label{equation:loss}
\end{align}
where $\epsilon \sim \mathcal{N}(0, I)$ is Gaussian noise, and the noise prediction network $\epsilon_\theta$ is modeled using a UNet architecture \cite{ronneberger2015u}, incorporating a cross-attention mechanism to integrate the text embedding \(\tau\).
% $\epsilon \sim \mathcal{N}(\bold{0}, \bold{I})$

Unlike the standard diffusion model training process, we first sample from a standard normal distribution and then use a pre-trained TTA model, i.e., Auffusion \cite{xue2024auffusion}, to generate the audio representation \(z_0\) from \(\epsilon\). The generated data and the corresponding noise are subsequently paired to retrain the LDM. 

\section{Experiments}

\subsection{Experimental Setup}

\subsubsection{Dataset}

Following previous works \cite{ghosal2023text, guan2024lafma}, we conduct our experiments using the AudioCaps (AC) \cite{kim2019audiocaps} dataset. It is a publicly available dataset that contains high-quality human-annotated captions. We first use Auffusion along with AudioCaps captions to generate $47,744$ ten-second audio samples for training. For evaluation, we use the test subset of AudioCaps, which is a widely recognized in-the-wild standard audio benchmark for TTA. Our test set consists of $928$ samples.
% a SOTA TTA model, 
% Unlike the standard training process of diffusion models,

% \subsubsection{Model and Training Details}
% \label{section:pdf_sanitise}
\subsubsection{Training Details and Baseline Models}
\label{section:pdf_sanitise}

We train only the parameters of the LDM while freezing all other modules in Auffusion. The original Auffusion model is built on Stable Diffusion v$1.5$\footnote{\url{https://huggingface.co/stable-diffusion-v1-5/stable-diffusion-v1-5}}, including its VAE and UNet. We fine-tune the UNet using our generated paired data \(\{(\epsilon, z_0)\}\) along with the corresponding text descriptions. Our proposed model is trained for $20,000$ iterations with a batch size of $128$ using the AdamW optimizer and a fixed learning rate of \(5 \times 10^{-6}\).

% , but it uses a custom-trained HiFi-GAN vocoder \cite{kong2020hifi}.
% We train the model using the AdamW optimizer with a fixed learning rate of \(5 \times 10^{-6}\), employing a constant learning rate scheduler. With a batch size of 128 and a total of 20,000 iterations, the model processes \( 128 \times 20,000 = 2,560,000 \) samples in total. In contrast, LAFMA processed a total of \( 45,122 \times 60 = 2,707,320 \) samples.
% Therefore, our total training cost is slightly lower than that of LAFMA.
% \subsubsection{Baseline Models}
% \label{section:pdf_sanitise}

In our paper, we compare AudioTurbo with several existing TTA models, including diffusion-based models such as AudioLDM2 \cite{liu2024audioldm}, Tango \cite{ghosal2023text}, and Auffusion\footnote{\url{https://huggingface.co/auffusion/auffusion}} \cite{xue2024auffusion}, as well as the flow-matching-based model LAFMA\footnote{\url{https://github.com/gwh22/LAFMA}} \cite{guan2024lafma}. Among these, LAFMA serves as our primary flow-matching-based acceleration model for comparison.

% as well as the autoregressive model AudioGen \cite{kreuk2022audiogen} and the flow-matching-based model LAFMA\footnote{\url{https://github.com/gwh22/LAFMA}} \cite{guan2024lafma}.

% AudioLDM \cite{liu2023audioldm}, 

\subsubsection{Evaluation Metrics}
\label{section:pdf_sanitise}

In this study, we evaluate our proposed TTA system, utilizing both objective metrics and subjective measures to assess the fidelity and diversity of the generated audio clips. For objective evaluation, we follow previous evaluation methods \cite{liu2023audioldm, ghosal2023text}, employing metrics such as Fréchet Distance (FD), Kullback-Leibler (KL) divergence, Inception Score (IS), and CLAP Score. 
% Similar to the Fréchet Inception Distance used in image synthesis, the FD score measures the distribution-level similarity between the embeddings of the generated audio clips and those of the target clips, without requiring paired reference audio samples. KL divergence is computed at the paired sample level based on the label distribution. IS is an effective metric for evaluating the quality and diversity of generated samples. All of these three metrics are based on the SOTA audio classifier PANNs \cite{kong2020panns}. In addition, we utilize a pre-trained CLAP model to measure the similarity between the generated audio and the text prompt, providing an evaluation of text-audio alignment. The evaluation package from this project\footnote{\url{https://github.com/haoheliu/audioldm_eval/}} is used to evaluate all these metrics.

% We use the evaluation package from this project\footnote{\url{https://github.com/haoheliu/audioldm_eval/}} to assess all these metrics.
% The evaluation package we use for all these metrics is included in the project\footnote{\url{https://github.com/haoheliu/audioldm_eval/}}.

% KL divergence is a reference-dependent metric that measures how much the distribution of generated audio differs from the reference audio distribution. Specifically, it utilizes a pre-trained classifier to estimate probabilities for both generated and reference samples, and then computes the KL divergence between these distributions.

% For subjective evaluation, we ask six evaluators to assess two aspects of 50 randomly selected audio samples—overall quality (OVL) and relevance to the input text (REL)—rating each on a scale from 1 to 100.
Following AudioLDM \cite{liu2023audioldm} and Tango \cite{ghosal2023text}, we conduct a subjective evaluation by asking six evaluators to assess two aspects of $50$ randomly selected audio samples: overall quality (OVL) and relevance to the input text (REL). Each aspect is rated on a scale from $1$ to $100$. The evaluators are colleagues from our research group with expertise in audio research.
% For subjective evaluation, we ask six evaluators to assess two aspects of $50$ randomly selected audio samples:

\subsection{Results and Analysis}
\label{section:multimedia}

\subsubsection{Main Results}
\label{section:supplementary}

We compare our proposed model, AudioTurbo, with other baselines on the AudioCaps test set. Our main results are presented in Table~\ref{tab:AudioCapsResults}. As the baseline systems typically achieve better performance with more inference steps, we set the inference steps to a standard value of $200$. However, the results of our proposed Audio-Turbo are achieved using step counts of $5$ and $10$.
% However, the best results of our proposed AudioTurbo shown in the table are achieved using only $10$ steps.

As shown in the table, AudioTurbo achieves strong performance across both objective and subjective metrics with only $10$ inference steps, whereas the other TTA baselines use $200$ steps. This demonstrates that, compared to other baseline models, AudioTurbo achieves better audio quality with fewer inference steps, highlighting the enhanced performance of our proposed method in both audio generation capability and inference efficiency. In addition, the proposed AudioTurbo achieves a CLAP score of $29.8$ and a REL of $85.58$ with $10$ sampling steps, outperforming other baselines by a large margin. This indicates that our proposed model can generate audio that is more relevant to the given textual description, demonstrating superior audio-text alignment capability. 

% Additionally, our model maintains strong performance even with just $5$ inference steps, without a significant decline.

Additionally, our model maintains strong performance even with just $5$ inference steps, with only a modest performance degradation. Overall, it still outperforms other models, except for the FD metric, where it scores $22.18$, slightly behind Tango’s $21.97$. To further assess the inference efficiency of the model, we report the Real-Time Factor (RTF), defined as the ratio between the total time required to synthesize an audio sample and the duration of the generated audio. All RTF measurements are conducted on the same NVIDIA A100 GPU to ensure a fair comparison. AudioTurbo attains the lowest real-time factor ($0.093$), yielding a 10×–20× acceleration compared to state-of-the-art diffusion-based TTA models.

Moreover, the proposed model is trained exclusively on text descriptions and the corresponding audio clips generated by the pre-trained teacher model, Auffusion. Integrating the rectified diffusion method for TTA does not require training the model from scratch and can leverage SOTA pre-trained models to directly generate audio-text pairs, moderately reducing training costs.
 
 % This provides further evidence of the effectiveness and superiority of the rectified diffusion method over flow-matching-based approaches in text-guided audio generation.

\subsubsection{Ablation Study on CFG Scale}
\label{section:repos}
% We found that a guidance scale of 1.5 delivers the best performance across nearly all metrics for AudioTurbo.
\begin{table}[ht!]
\centering
\small
\caption{Impact on objective evaluation metrics with varying levels of classifier-free guidance.}
\resizebox{8cm}{!}{
\begin{tabular}{c|cc|cccc}
\toprule
\multirow{1}{*}{\textbf{Model}}
& \textbf{Steps} & \textbf{Guidance} & \textbf{FD}~$\downarrow$ & \textbf{KL}~$\downarrow$ & \textbf{IS}~$\uparrow$ & 
\textbf{CLAP(\%)}~$\uparrow$ \\
\midrule
\multirow{5}{*}{AudioTurbo} & \multirow{5}{*}{$25$} 
& $1.0$ & $23.20$ & $1.35$ & $8.95$ & $28.9$ \\
& & $1.5$ & $\mathbf{21.69}$ & $\mathbf{1.31}$ & $9.74$ & $\mathbf{29.6}$ \\
& & $2.5$ & $21.98$ & $1.36$ & $\mathbf{9.86}$ & $28.8$ \\
% & & $2.5$  & $\mathbf{37.46}$ & $\mathbf{1.40}$ & $\mathbf{1.85}$  \\
& & $5.0$ & $26.18$ & $1.51$ & $9.16$ & $25.1$ \\
& & $7.5$ & $32.49$ & $1.72$ & $7.56$ & $20.4$ \\
\bottomrule
\end{tabular}
}
\label{tab:Guidance}
\end{table}

Since the classifier-free guidance scale plays an important role in generation quality and sampling diversity \cite{liu2023audioldm}, we investigated the impact of the CFG scale on AudioTurbo with the inference step fixed at $25$. The results are presented in Table~\ref{tab:Guidance}. The first row represents a guidance scale of $1.0$, meaning that classifier-free guidance is not applied during inference. The result of this configuration is not particularly impressive. This highlights the importance of incorporating classifier-free guidance. We achieve the best performance across nearly all metrics for AudioTurbo with a guidance scale of $1.5$, while the metrics deteriorate as the guidance scale increases further. These findings underscore the critical role of proper guidance scale selection, as an appropriate balance must be achieved to optimize performance across all objective metrics.
% Adjusting the guidance scale carefully allows us to improve the overall effectiveness of our approach. Therefore, in the other experiments of this paper, we fix the guidance scale of AudioTurbo at $1.5$. 

% Our findings shed light on how the CFG scale impacts the performance of our proposed TTA model, AudioTurbo. Adjusting the guidance scale carefully allows us to enhance the alignment between generated audio samples and the conditioning information while maintaining an acceptable level of diversity, thereby improving the overall effectiveness of our approach.

\begin{table}[ht!]
\centering
\small
% \normalsize
\caption{Objective metric values for LAFMA, Auffusion, and AudioTurbo models at different inference steps.}
\resizebox{8cm}{!}{
\begin{tabular}{cc|ccccc}
\toprule
\multirow{2}{*}{\textbf{Metrics}} & \multirow{2}{*}{\textbf{Model}}  & \multicolumn{5}{c}{\textbf{Inference Steps}} \\
& & 3 & 5 & 10 & 25 & 200  \\
\midrule
\multirow{3}{*}{FD~$\downarrow$} & LAFMA & $91.53$ & $53.46$ & $31.26$ & $27.37$ & $29.59$ \\
% & AudioTurbo & $\mathbf{29.96}$ & $\mathbf{22.18}$ & $\mathbf{20.65}$ & $\underline{21.69}$ & $\underline{21.75}$ \\
& Auffusion & $55.66$ & $36.53$ & $25.17$ & $\mathbf{21.21}$ & $\mathbf{21.31}$ \\
& AudioTurbo & $\mathbf{29.96}$ & $\mathbf{22.18}$ & $\mathbf{20.65}$ & $\underline{21.69}$ & $\underline{21.75}$ \\
\midrule

\multirow{3}{*}{KL~$\downarrow$} & LAFMA & $3.30$ & $2.13$ & $1.61$ & $1.58$ & $1.56$ \\
% & AudioTurbo & $\mathbf{1.52}$ & $\mathbf{1.30}$ & $\mathbf{1.29}$ & $\underline{1.31}$ & $\underline{1.32}$ \\
& Auffusion & $2.75$ & $1.98$ & $1.65$ & $\mathbf{1.27}$ & $\mathbf{1.30}$ \\
& AudioTurbo & $\mathbf{1.52}$ & $\mathbf{1.30}$ & $\mathbf{1.29}$ & $\underline{1.31}$ & $\underline{1.32}$ \\
\midrule

\multirow{3}{*}{CLAP (\%)~$\uparrow$} & LAFMA & $-1.8$ & $9.7$ & $19.8$ & $22.9$ & $22.9$ \\
% & AudioTurbo & $\mathbf{25.4
% }$ & $\mathbf{29.2}$ & $\mathbf{29.8}$ & $\underline{29.6}$ & $\underline{29.6}$ \\
& Auffusion & $8.4$ & $15.5$ & $21.6$ & $\mathbf{30.2}$ & $\mathbf{30.5}$ \\
& AudioTurbo & $\mathbf{25.4
}$ & $\mathbf{29.2}$ & $\mathbf{29.8}$ & $\underline{29.6}$ & $\underline{29.6}$ \\
\bottomrule
\end{tabular}
}
\label{tab:compare}
\end{table}

\subsubsection{Comparative Study of TTA Models with Varying Inference Steps}
\label{section:repos}

We report the results of a comparative study on TTA models with varying inference steps in Table~\ref{tab:compare}. It can be observed that increasing the number of sampling steps generally improves sample quality in the low-step regime. However, once a sufficient number of steps is reached, the performance tends to saturate and may exhibit minor fluctuations. With the same number of inference steps, our model achieves the best or near-best performance, with its advantage being especially prominent in the low-step regime.

Furthermore, compared to the flow-matching-based TTA acceleration model (LAFMA), our model achieves comparable performance in just three steps, matching LAFMA’s optimal performance, which is achieved at $25$ steps, while significantly improving inference efficiency.
% Furthermore, compared to the flow-matching-based TTA acceleration model (LAFMA), our model achieves comparable performance in just three steps, matching LAFMA's performance at $25$ steps while significantly improving inference efficiency. 
% In the future, we aim to further enhance model performance and achieve one-step generation using distillation techniques.

\subsubsection{Generalizability of Our Method}
\label{section:generality}

\begin{table}[ht!]
\caption{Evaluation of our method’s generalizability to AudioLDM on the AudioCaps test set.}
\centering
\small
\resizebox{0.48\textwidth}{!}
{%
\begin{tabular}{cc|cccc}
\toprule

 % \multirow{1}{*}{\textbf{Model}} &\multicolumn{1}{c}{\textbf{Steps}}  &\multicolumn{1}{c}{\textbf{FD}$\downarrow$} 
 %  & \multicolumn{1}{c}{\textbf{KL}$\downarrow$} & \multicolumn{1}{c}{\textbf{IS} $\uparrow$} & \multicolumn{1}{c}{\textbf{CLAP (\%)} $\uparrow$}  \\
\textbf{Model} & \multicolumn{1}{c|}{\textbf{Steps}} &
\multicolumn{1}{c}{\textbf{FD}$\downarrow$} &
\multicolumn{1}{c}{\textbf{KL}$\downarrow$} &
\multicolumn{1}{c}{\textbf{IS}$\uparrow$} &
\multicolumn{1}{c}{\textbf{CLAP (\%) }$\uparrow$} \\

\midrule
AudioLDM & $200$&$35.04$ & $1.81$ & $7.49$ & $18.4$  \\

AudioLDM & $25$& $38.3$&$1.95$ & $7.15$ & $15.6$  \\

AudioLDM & $10$&$47.96$ & $2.16$ & $5.93$ & $12.5$  \\

AudioLDM & $5$&$63.46$ & $2.68$ & $4.23$ & $7.8$  \\
\midrule
AudioLDM-Turbo & $25$& $32.34$ & $1.69$ & $7.24$ & $17.4$ \\

AudioLDM-Turbo & $10$& $31.99$ & $1.70$ & $7.28$ & $17.2$ \\

AudioLDM-Turbo & $5$& $33.31$ & $1.77$ & $7.03$ & $16.9$  \\

% AudioLDM-Turbo & $3$& $40.37$ & $2.25$ & $6.52$ & $14.8$  \\
% \midrule
\bottomrule
\end{tabular}

}

\label{tab:general}
% \vspace{-0.4cm} % 减少0.5厘米的垂直空间
\end{table}

Our framework can be applied not only to Auffusion but also to other diffusion-based TTA models to improve inference speed while maintaining comparable generation quality. To further demonstrate the generalizability of our method, we conducted additional experiments on AudioLDM, and refer to the accelerated model as AudioLDM-Turbo. The model’s performance is evaluated on the AudioCaps test set.

As shown in Table~\ref{tab:general}, AudioLDM exhibits a significant performance drop as the number of inference steps decreases. When reduced to $25$ steps, its generation quality becomes marginal and is even surpassed by AudioLDM-Turbo operating with only $5$ steps. However, for AudioLDM-Turbo, reducing the number of inference steps does not lead to a noticeable performance degradation. The model maintains strong performance even with fewer inference steps, as noise and data samples are deterministically paired during training. This allows the model to generate consistent predictions along a single trajectory, thereby enabling effective acceleration.

\section{Conclusions}

In this paper, we have presented a rectified diffusion-based post-training distillation framework for text-to-audio generation, designed to leverage the strengths of the SOTA pre-trained TTA models while enhancing inference efficiency. Experiments show that by combining pre-trained TTA models with rectified diffusion, AudioTurbo achieves competitive performance across both objective and subjective evaluations while substantially reducing the number of inference steps. Moreover, it achieves comparable generation quality with only $3$ steps compared to a flow-matching-based acceleration model with $25$ steps, effectively reducing computational overhead.

\section*{Acknowledgment}
This work was supported in part by the Engineering and Physical Sciences Research Council (EPSRC) under Grant EP/T019751/1, funded by a PhD scholarship from the Doctoral College and the Centre for Vision, Speech, and Signal Processing (CVSSP) at the University of Surrey. For the purpose of open access, the authors have applied a Creative Commons Attribution (CC BY) licence to any Author Accepted Manuscript version arising.

% \begin{thebibliography}{1}

% \bibitem{1}
% G.~Eason, B.~Noble, and I.~N.~Sneddon, ``On certain integrals of
% Lipschitz-Hankel type involving products of Bessel functions,''
% \emph{Phil. Trans. Roy. Soc. London,} vol. A247, pp. 529-551, April
% 1955.

% \bibitem{2}
% J.~Clerk~Maxwell, \emph{A Treatise on Electricity and Magnetism,}
% 3$^{\rm rd}$ ed., vol. 2. Oxford: Clarendon, 1892, pp.68-73.

% \bibitem{3}
% I.~S.~Jacobs and C.~P.~Bean, ``Fine particles, thin films and exchange
% anisotropy,'' in \emph{Magnetism,} vol. III, G.T. Rado and H. Suhl,
% Eds. New York: Academic, 1963, pp. 271-350.

% \bibitem{4}
% K.~Elissa, ``Title of paper if known,'' unpublished.

% \bibitem{5}
% R.~Nicole, ``Title of paper with only first word capitalized,''
% \emph{J. Name Stand. Abbrev.,} in press.

% \bibitem{6}
% Y.~Yorozu, M.~Hirano, K.~Oka, and Y.~Tagawa, ``Electron spectroscopy
% studies on magneto-optical media and plastic substrate interface,''
% \emph{APSIPA Transl. J. Magn. Japan,} vol. 2, pp. 740-741, August 1987
% [\emph{Digests 9$^{\rm th}$ Annual Conf. Magnetics Japan,} p. 301,
% 1982].

% \bibitem{7}
% M.~Young, \emph{The Technical Writer's Handbook.} Mill Valley, CA:
% University Science, 1989.

% \end{thebibliography}

\printbibliography

@article{zhu2026audio,
  title={Audio ControlNet for Fine-Grained Audio Generation and Editing},
  author={Zhu, Haina and Xiao, Yao and Li, Xiquan and Ma, Ziyang and Yu, Jianwei and Zhang, Bowen and Yang, Mingqi and Chen, Xie},
  journal={arXiv preprint arXiv:2602.04680},
  year={2026}
}

@article{xie2024picoaudio,
  title={Picoaudio: Enabling precise timestamp and frequency controllability of audio events in text-to-audio generation},
  author={Xie, Zeyu and Xu, Xuenan and Wu, Zhizheng and Wu, Mengyue},
  journal={arXiv preprint arXiv:2407.02869},
  year={2024}
}

@article{yuan2025dreamaudio,
  title={DreamAudio: Customized Text-to-Audio Generation with Diffusion Models},
  author={Yuan, Yi and Liu, Xubo and Liu, Haohe and Kang, Xiyuan and Chen, Zhuo and Wang, Yuxuan and Plumbley, Mark D and Wang, Wenwu},
  journal={arXiv preprint arXiv:2509.06027},
  year={2025}
}

@inproceedings{xu2024prompt,
  title={Prompt-guided Precise Audio Editing with Diffusion Models},
  author={Xu, Manjie and Li, Chenxing and Zhang, Duzhen and Su, Dan and Liang, Wei and Yu, Dong},
  booktitle={International Conference on Machine Learning},
  pages={55126--55143},
  year={2024},
  organization={PMLR}
}

@inproceedings{hou2025editing,
  title={Editing music with melody and text: Using controlnet for diffusion transformer},
  author={Hou, Siyuan and Liu, Shansong and Yuan, Ruibin and Xue, Wei and Shan, Ying and Zhao, Mangsuo and Zhang, Chao},
  booktitle={ICASSP 2025-2025 IEEE International Conference on Acoustics, Speech and Signal Processing (ICASSP)},
  pages={1--5},
  year={2025},
  organization={IEEE}
}

@inproceedings{liu2023audioldm,
  title={{AudioLDM}: text-to-audio generation with latent diffusion models},
  author={Liu, Haohe and Chen, Zehua and Yuan, Yi and Mei, Xinhao and Liu, Xubo and Mandic, Danilo and Wang, Wenwu and Plumbley, Mark D},
  booktitle={Proceedings of the 40th International Conference on Machine Learning},
  pages={21450--21474},
  year={2023}
}

@article{yang2023diffsound,
  title={{Diffsound}: Discrete diffusion model for text-to-sound generation},
  author={Yang, Dongchao and Yu, Jianwei and Wang, Helin and Wang, Wen and Weng, Chao and Zou, Yuexian and Yu, Dong},
  journal={IEEE/ACM Transactions on Audio, Speech, and Language Processing},
  volume={31},
  pages={1720--1733},
  year={2023},
  publisher={IEEE}
}

@inproceedings{ghosal2023text,
author = {Ghosal, Deepanway and Majumder, Navonil and Mehrish, Ambuj and Poria, Soujanya},
title = {Text-to-Audio Generation using Instruction Guided Latent Diffusion Model},
year = {2023},
booktitle = {Proceedings of the 31st ACM International Conference on Multimedia},
pages = {3590--3598}
}

@inproceedings{kreukaudiogen,
  title={{AudioGen}: Textually Guided Audio Generation},
  author={Kreuk, Felix and Synnaeve, Gabriel and Polyak, Adam and Singer, Uriel and D{\'e}fossez, Alexandre and Copet, Jade and Parikh, Devi and Taigman, Yaniv and Adi, Yossi},
  booktitle={The Eleventh International Conference on Learning Representations}
}

@article{yang2023uniaudio,
  title={{UniAudio}: An audio foundation model toward universal audio generation},
  author={Yang, Dongchao and Tian, Jinchuan and Tan, Xu and Huang, Rongjie and Liu, Songxiang and Chang, Xuankai and Shi, Jiatong and Zhao, Sheng and Bian, Jiang and Wu, Xixin and others},
  journal={arXiv preprint arXiv:2310.00704},
  year={2023}
}

@inproceedings{rombach2022high,
  title={High-resolution image synthesis with latent diffusion models},
  author={Rombach, Robin and Blattmann, Andreas and Lorenz, Dominik and Esser, Patrick and Ommer, Bj{\"o}rn},
  booktitle={Proceedings of the IEEE/CVF Conference on Computer Vision and Pattern Recognition},
  pages={10684--10695},
  year={2022}
}

@article{xue2024auffusion,
author = {Xue, Jinlong and Deng, Yayue and Gao, Yingming and Li, Ya},
title = {Auffusion: Leveraging the Power of Diffusion and Large Language Models for Text-to-Audio Generation},
year = {2024},
issue_date = {2024},
publisher = {IEEE Press},
volume = {32},
issn = {2329-9290},
journal = {IEEE/ACM Transactions on Audio, Speech, and Language Processing},
pages = {4700--4712},
numpages = {13}
}

@article{liu2024audioldm,
  title={{AudioLDM 2}: Learning holistic audio generation with self-supervised pretraining},
  author={Liu, Haohe and Yuan, Yi and Liu, Xubo and Mei, Xinhao and Kong, Qiuqiang and Tian, Qiao and Wang, Yuping and Wang, Wenwu and Wang, Yuxuan and Plumbley, Mark D},
  journal={IEEE/ACM Transactions on Audio, Speech, and Language Processing},
  year={2024},
  publisher={IEEE}
}

@inproceedings{huang2023make,
  title={{Make-an-Audio}: Text-to-audio generation with prompt-enhanced diffusion models},
  author={Huang, Rongjie and Huang, Jiawei and Yang, Dongchao and Ren, Yi and Liu, Luping and Li, Mingze and Ye, Zhenhui and Liu, Jinglin and Yin, Xiang and Zhao, Zhou},
  booktitle={International Conference on Machine Learning},
  pages={13916--13932},
  year={2023},
  organization={PMLR}
}

@article{huang2023make2,
  title={{Make-an-Audio 2}: Temporal-enhanced text-to-audio generation},
  author={Huang, Jiawei and Ren, Yi and Huang, Rongjie and Yang, Dongchao and Ye, Zhenhui and Zhang, Chen and Liu, Jinglin and Yin, Xiang and Ma, Zejun and Zhao, Zhou},
  journal={arXiv preprint arXiv:2305.18474},
  year={2023}
}

@inproceedings{mehta2024matcha,
  title={{Matcha-TTS}: A fast {TTS} architecture with conditional flow matching},
  author={Mehta, Shivam and Tu, Ruibo and Beskow, Jonas and Sz{\'e}kely, {\'E}va and Henter, Gustav Eje},
  booktitle={International Conference on Acoustics, Speech and Signal Processing (ICASSP)},
  pages={11341--11345},
  year={2024},
  organization={IEEE}
}

@inproceedings{ye2024flashspeech,
  title={{FlashSpeech}: Efficient zero-shot speech synthesis},
  author={Ye, Zhen and Ju, Zeqian and Liu, Haohe and Tan, Xu and Chen, Jianyi and Lu, Yiwen and Sun, Peiwen and Pan, Jiahao and Bian, Weizhen and He, Shulin and others},
  booktitle={Proceedings of the 32nd ACM International Conference on Multimedia},
  pages={6998--7007},
  year={2024}
}

@article{guan2024lafma,
  title={{LAFMA}: A Latent Flow Matching Model for Text-to-Audio Generation},
  author={Guan, Wenhao and Wang, Kaidi and Zhou, Wangjin and Wang, Yang and Deng, Feng and Wang, Hui and Li, Lin and Hong, Qingyang and Qin, Yong},
  journal={arXiv preprint arXiv:2406.08203},
  year={2024}
}

@article{wang2024rectified,
  title={Rectified diffusion: Straightness is not your need in rectified flow},
  author={Wang, Fu-Yun and Yang, Ling and Huang, Zhaoyang and Wang, Mengdi and Li, Hongsheng},
  journal={arXiv preprint arXiv:2410.07303},
  year={2024}
}

@inproceedings{
  song2020score,
  title={Score-Based Generative Modeling through Stochastic Differential Equations},
  author={Yang Song and Jascha Sohl-Dickstein and Diederik P Kingma and Abhishek Kumar and Stefano Ermon and Ben Poole},
  booktitle={International Conference on Learning Representations},
  year={2021},
}

@article{lu2022dpm,
  title={{DPM-Solver}: A fast {ODE} solver for diffusion probabilistic model sampling in around 10 steps},
  author={Lu, Cheng and Zhou, Yuhao and Bao, Fan and Chen, Jianfei and Li, Chongxuan and Zhu, Jun},
  journal={Advances in Neural Information Processing Systems},
  volume={35},
  pages={5775--5787},
  year={2022}
}

@inproceedings{wu2023large,
  title={Large-scale contrastive language-audio pretraining with feature fusion and keyword-to-caption augmentation},
  author={Wu, Yusong and Chen, Ke and Zhang, Tianyu and Hui, Yuchen and Berg-Kirkpatrick, Taylor and Dubnov, Shlomo},
  booktitle={International Conference on Acoustics, Speech and Signal Processing (ICASSP)},
  pages={1--5},
  year={2023},
  organization={IEEE}
}

@inproceedings{radford2021learning,
  title={Learning transferable visual models from natural language supervision},
  author={Radford, Alec and Kim, Jong Wook and Hallacy, Chris and Ramesh, Aditya and Goh, Gabriel and Agarwal, Sandhini and Sastry, Girish and Askell, Amanda and Mishkin, Pamela and Clark, Jack and others},
  booktitle={International Conference on Machine Learning},
  pages={8748--8763},
  year={2021},
  organization={PMLR}
}

@inproceedings{kim2019audiocaps,
  title={{AudioCaps}: Generating captions for audios in the wild},
  author={Kim, Chris Dongjoo and Kim, Byeongchang and Lee, Hyunmin and Kim, Gunhee},
  booktitle={Proceedings of the 2019 Conference of the North American Chapter of the Association for Computational Linguistics: Human Language Technologies, Volume 1 (Long and Short Papers)},
  pages={119--132},
  year={2019}
}

@article{van2017neural,
  title={Neural discrete representation learning},
  author={Van Den Oord, Aaron and Vinyals, Oriol and others},
  journal={Advances in Neural Information Processing Systems},
  volume={30},
  year={2017}
}

@inproceedings{ronneberger2015u,
  title={{U-Net}: Convolutional networks for biomedical image segmentation},
  author={Ronneberger, Olaf and Fischer, Philipp and Brox, Thomas},
  booktitle={Medical Image Computing and Computer-Assisted Intervention},
  pages={234--241},
  year={2015},
  organization={Springer}
}

@inproceedings{bai2024consistencytta,
  title     = {{ConsistencyTTA}: Accelerating Diffusion-Based Text-to-Audio Generation with Consistency Distillation},
  author    = {Bai, Yatong and Dang, Trung and Tran, Dung N. and Koishida, Kazuhito and Sojoudi, Somayeh},
  booktitle = {Proceedings of Interspeech},
  year      = {2024}
}

@inproceedings{saito2024soundctm,
  title     = {{SoundCTM}: Uniting Score-Based and Consistency Models for Text-to-Sound Generation},
  author    = {Saito, Koichi and Kim, Dongjun and Shibuya, Takashi and Lai, Chieh-Hsin and Zhong, Zhi and Takida, Yuhta and Mitsufuji, Yuki},
  booktitle = {Audio Imagination: NeurIPS 2024 Workshop on AI-Driven Speech, Music, and Sound Generation},
  year      = {2024}
}

@inproceedings{liu2024audiolcm,
  title     = {{AudioLCM}: Efficient and High-Quality Text-to-Audio Generation with Minimal Inference Steps},
  author    = {Liu, Huadai and Huang, Rongjie and Liu, Yang and Cao, Hengyuan and Wang, Jialei and Cheng, Xize and Zheng, Siqi and Zhao, Zhou},
  booktitle = {Proceedings of the 32nd ACM International Conference on Multimedia},
  pages     = {7008--7017},
  year      = {2024},
  doi       = {10.1145/3664647.3681072}
}

@inproceedings{liu2025flashaudio,
  title     = {{FlashAudio}: Rectified Flow for Fast and High-Fidelity Text-to-Audio Generation},
  author    = {Liu, Huadai and Wang, Jialei and Huang, Rongjie and Liu, Yang and Lu, Heng and Zhao, Zhou and Xue, Wei},
  booktitle = {Proceedings of the 63rd Annual Meeting of the Association for Computational Linguistics (Volume 1: Long Papers)},
  pages     = {13694--13710},
  year      = {2025},
  doi       = {10.18653/v1/2025.acl-long.673}
}

\end{document}